\documentclass[12pt,preprint]{aastex}

\accepted{}

\def\gsim{\;\rlap{\lower 2.5pt
 \hbox{$\sim$}}\raise 1.5pt\hbox{$>$}\;}
\def\lsim{\;\rlap{\lower 2.5pt
   \hbox{$\sim$}}\raise 1.5pt\hbox{$<$}\;}

\slugcomment{30 August 2005 Revision; AJ accepted}

\received{2005 July 7}
\begin{document}

\title{A Snapshot Survey for Gravitational Lenses Among $z \ge 4.0$
Quasars: II. Constraints on the $4.0<z<4.5$ Quasar
Population\footnote{Based on observations made with the NASA/ESA
Hubble Space Telescope, obtained at the Space Telescope Science
Institute, which is operated by the Association of Universities for
Research in Astronomy, Inc., under NASA contract NAS 5-26555. These
observations are associated with program \#9472.}}

\author{
Gordon T. Richards,\altaffilmark{2}
Zolt\'{a}n Haiman,\altaffilmark{3}
Bartosz Pindor,\altaffilmark{4}
Michael A. Strauss,\altaffilmark{2}
Xiaohui Fan,\altaffilmark{5}
Daniel Eisenstein,\altaffilmark{5}
Donald P. Schneider,\altaffilmark{6}
Neta A. Bahcall,\altaffilmark{2}
J. Brinkmann,\altaffilmark{7}
and Masataka Fukugita\altaffilmark{8}
}

\altaffiltext{2}{Princeton University Observatory, Peyton Hall, Princeton, NJ 08544.}
\altaffiltext{3}{Department of Astronomy, Columbia University, 550 West 120th Street, New York, NY 10027.}
\altaffiltext{4}{Department of Astronomy, University of Toronto, 60 St. George Street, Toronto M5S 3H8, Canada.}
\altaffiltext{5}{Steward Observatory, University of Arizona, 933 North Cherry Avenue, Tucson, AZ 85721.}
\altaffiltext{6}{Department of Astronomy and Astrophysics, The Pennsylvania State University, 525 Davey Laboratory, University Park, PA 16802.}
\altaffiltext{7}{Apache Point Observatory, P.O. Box 59, Sunspot, NM 88349.}
\altaffiltext{8}{ICRR/Institute for Cosmic Ray Research, University of Tokyo, 5-1-5 Kashiwa, Kashiwa City, Chiba 277-8582, Japan.}

\begin{abstract}

We report on $i$-band snapshot observations of 157 Sloan Digital Sky
Survey (SDSS) quasars at $4<z<5.4$ using the Advanced Camera for
Surveys on the {\em Hubble Space Telescope (HST)} to search for
evidence of gravitational lensing of these sources.  None of the
quasars appear to be strongly lensed and multiply imaged at the
angular resolution ($\sim0\farcs1$) and sensitivity of {\it HST}.  The
non-detection of strong lensing in these systems constrains the
$z=4$--5 luminosity function to an intrinsic slope of $\beta>-3.8$
$(3\sigma)$, assuming a break in the quasar luminosity function at
$M_{1450}^{\star}\sim-24.5$.  This constraint is considerably stronger
than the limit of $\beta>-4.63$ obtained from the absence of lensing
in four $z>5.7$ quasars.  Such constraints are important for our
understanding of the true space density of high-redshift quasars and
the ionization state of the early universe.

\end{abstract}


\section{Introduction}

Investigations of multiply--imaged high-redshift quasars are important
for our basic understanding of the formation and growth of
supermassive black holes in galactic centers
\markcite{tur91,hl01}({Turner} 1991; {Haiman} \& {Loeb} 2001) and the
ionization state of the universe as a function of time
\markcite{mhr99,wl03}({Madau}, {Haardt}, \& {Rees} 1999; {Wyithe} \&
{Loeb} 2003a).  Gravitational lensing changes the apparent flux coming
from a quasar and thus changes our interpretation of flux-dependent
properties; the discovery that a given quasar is gravitationally
lensed means that the naively estimated luminosity is too high.  Thus,
gravitational lensing would modify the expected size of ionized
(\ion{H}{2}) regions around individual quasars \markcite{ch00}({Cen}
\& {Haiman} 2000) and would weaken the lower limits on the neutral
fraction in the IGM as inferred from the size of \ion{H}{2} regions
\markcite{wl04,mh04}({Wyithe} \& {Loeb} 2004; {Mesinger} \& {Haiman}
2004).  This paper concentrates on the effect that gravitational
lensing has on the apparent shape of the quasar luminosity function
\markcite{chs02,wl02,fss+03}({Comerford}, {Haiman}, \& {Schaye} 2002;
{Wyithe} \& {Loeb} 2002; {Fan} {et~al.} 2003).  In a previous paper
\markcite{rsp+04}({Richards} {et~al.} 2004, hereafter Paper~I) we
investigated whether the known $z\sim6$ quasars are gravitationally
lensed and discussed how the lack of lenses in this sample affects our
understanding of the growth of black holes in the early universe
\markcite{hl01}(e.g., {Haiman} \& {Loeb} 2001) and the ionization
history of the universe at the end of the reionization period
\markcite{fns+02}(e.g., {Fan} {et~al.} 2002).  In this paper, we
examine the constraints that can be placed on the intrinsic slope of
the quasar luminosity function (QLF) at $z\sim$4--5 from a search for
gravitational lenses in a sample of $z\sim$4--5 quasars
\markcite{fsr+01,afr+01,sfs+01,shr+05}({Fan} {et~al.} 2001a;
{Anderson} {et~al.} 2001; {Schneider} {et~al.} 2001; {Schneider et
al.} 2005) in the Sloan Digital Sky Survey (SDSS;
\markcite{yaa+00}{York} {et~al.} 2000).

Much theoretical effort has been devoted to placing constraints on the
slope of the quasar luminosity function from the fraction of lenses
found amongst high-$z$ quasars \markcite{chs02,wl02}(e.g., {Comerford}
{et~al.} 2002; {Wyithe} \& {Loeb} 2002).  \markcite{chs02}{Comerford}
{et~al.} (2002) showed that modest constraints could be obtained from
a sample of $\sim20$ $z\sim6$ quasars from ground-based imaging (i.e.,
sensitive to $\ge1\arcsec$ splittings), and that similar limits could
be derived from {\em Hubble Space Telescope (HST)} resolution imaging
of the (then) four known $z\sim6$ quasars.  Paper I presented {\em
HST} imaging showing that none of those four quasars are strongly
gravitationally lensed and derived a limit on the slope of the bright
end quasar luminosity function of $\beta>-4.63 (3\sigma)$.
\markcite{wyi04}{Wyithe} (2004) found even stronger constraints for
this sample by including the observation that two quasars appear to be
lensed by foreground galaxies but are not multiply imaged.

These limits come from magnification bias
\markcite{tog84,nar89}({Turner}, {Ostriker}, \& {Gott} 1984; {Narayan}
1989), which would be strong if the slope of the QLF were steep and
there were many quasars with intrinsic luminosities below the
detection threshold.  As discussed in Paper I, the expected fraction
of multiply-imaged quasars at a given redshift depends both on the
cosmological model and on the luminosity function of quasars.  In the
WMAP cosmology (\markcite{svp+03}{Spergel} {et~al.} 2003, using a halo
distribution taken from the large N-body simulations of
\markcite{jfw+01}{Jenkins} {et~al.} 2001), the fraction of random
lines of sight out to $z=4$ that produce multiple images at all
splitting angles is of order 0.2\%; this fraction rises to $\sim0.4$\%
at $z=6$.  If the intrinsic (unlensed) QLF has a break in slope, then,
in general, the fraction of lensed quasars would increase for apparent
fluxes above the break.  In a flux-limited survey, there is a strong
correlation between luminosity and redshift.  If the true QLF
possesses a break which moves to fainter luminosities at higher
redshift, then the most distant quasars (which also look through the
longest path length) are expected to be the most likely to be lensed.
More explicitly, in the event that no lensing is observed, Paper I
showed that the tightest constraints on the QLF slope come from
$z\sim6$ quasars, and also that roughly seven $z\sim4$ quasars have
the same statistical power as a single $z\sim6$ quasar.  Herein we
present the results for a sample of 157 $z\sim4$ quasars from the SDSS
that were imaged with {\em HST}; we find that none of these sources is
lensed.  As predicted, these quasars provide a constraint on the slope
of the QLF that is roughly equivalent to {\em HST} imaging of 22
$z\sim6$ quasars, limiting the bright end slope of the $z=4$--5 QLF to
$\beta>-3.8 (3\sigma)$.

Section~2 describes the sample and the data.  In \S~3 we discuss the
constraints on the QLF that can be derived from the data.  We
summarize in \S~4.  Throughout this paper, we adopt the WMAP cosmology
with $\Omega_m=0.3$, $\Omega_\Lambda=0.7$, $H_0=70\,{\rm
km\,s^{-1}\,Mpc^{-1}}$, an rms mass fluctuation within a sphere of
radius $8 \; h^{-1}$ Mpc of $\sigma_8=0.9$, and power--law index
$n=0.99$ for the power spectrum of density fluctuations
\markcite{svp+03}({Spergel} {et~al.} 2003).  We also adopt the
cosmological transfer function from \markcite{eh99}{Eisenstein} \&
{Hu} (1999).  Conversions between $M_B$ and $M_{1450}$ assume $M_B =
M_{1450} - 0.48$ \markcite{ssg95}({Schmidt}, {Schneider}, \& {Gunn}
1995) with spectral index
$\alpha_{\nu}=-0.5\;(f_{\nu}\propto\nu^{\alpha_{\nu}})$.

\section{The Data}

\subsection{Observations and Data Processing}

There were 281 SDSS quasars with $z \ge 4.0$ as of January 2002
\markcite{fss+99,fss+00,zts+00,fsr+01,sfs+01,afr+01}({Fan} {et~al.} 1999, 2000; {Zheng} {et~al.} 2000; {Fan} {et~al.} 2001a; {Schneider} {et~al.} 2001; {Anderson} {et~al.} 2001), when the sample was
defined.  Most of these were included in SDSS spectroscopy, but
several were discovered as part of early follow-up spectroscopy on the
ARC 3.5m telescope and other telescopes \markcite{fss+99}(e.g., {Fan} {et~al.} 1999),
including some that have not yet been published.  We were granted the
opportunity to observe 250 of these in HST snapshot mode; we did so by
including in the sample all quasars with $z \ge 4.6$, and all quasars
with $4.0 < z < 4.6$ with $i < 20.3$.

Snapshot observations are carried out with the understanding that not
all objects will be observed.  In this context, we were able to give
objects priorities for being observed.  We put the 48 objects with
absolute 1450\AA\ magnitude less than $-29$, and/or with $z>5$ at
highest priority, and those with $M_{1450} > -27.95$ at lowest
priority.  All remaining objects, and those with $4.7 < z < 5$, were
placed at medium priority.  At the end, 161 objects were observed (of
which four were presented in Paper~I): 48/48 of the high-priority
objects, 97/154 of the medium priority objects, and 16/48 of the
low-priority objects.

The four $z>5.7$ quasars included in our {\em HST} snapshot program
were presented in Paper I.  In this paper, we describe the results of
imaging of 157 SDSS quasars with $4<z<5.4$; see Table~\ref{tab:tab1}.
For the sake of completeness, we also tabulate those 89 sources that
were included in our sample, but were never observed by {\em HST}, see
Table~\ref{tab:tab2}.  Seven objects in Table~1 and five objects in
Table~2 are previously unpublished.

Images of the 157 quasars were acquired with the High Resolution
Camera (HRC) on the Advanced Camera for Surveys (ACS).  Observations
were taken in the SDSS $i$-band (F775W).  The Wide-Field Camera on ACS
has higher sensitivity in $i$ than does the HRC, but has substantially
higher overhead and is mildly undersampled.  The exposure times were
640 seconds for each object, 320 seconds in each of two exposures to
help in cosmic ray rejection.

The data processing was discussed in detail in Paper I, but we briefly
review the process here.  The raw images were calibrated by the CALACS
package in IRAF\footnote{IRAF is distributed by the National Optical
Astronomy Observatories, which are operated by the Association of
Universities for Research in Astronomy, Inc., under cooperative
agreement with the National Science Foundation.} as part of
on-the-fly-reprocessing (OTFR) at the time of download.  The images
that we present are the ``cosmic ray rejected'' (CRJ) images that are
output by the OTFR algorithms at STScI.  The CRJ files have all been
reduced in the standard manner, including having been overscan-, bias-
and dark-corrected, flat-fielded and photometrically calibrated, in
addition to having bad pixels masked and cosmic rays removed (see the
ACS
manual\footnote{http://www.stsci.edu/hst/acs/documents/handbooks/cycle12/cover.html}
for more details).

\subsection{Subtracting the Point Source and Looking for Multiple Images}

{\em HST} resolution data are needed for this project as the median
expected splitting of gravitationally lensed quasars is predicted (and
observed) to be somewhat less than $1\arcsec$
\markcite{tog84,hk87,bwj+03}({Turner} {et~al.} 1984; {Hinshaw} \& {Krauss} 1987; {Browne} {et~al.} 2003) and the SDSS images themselves have
point-spread-function (PSF) widths of order $1\arcsec - 1\farcs8$
\markcite{aaa+03}({Abazajian} {et~al.} 2003) with an image scale of $0\farcs396\,{\rm pixel}^{-1}$.
Our ACS images, on the other hand, have an image scale of
$0\farcs025\,{\rm pixel}^{-1}$ and the point spread function is narrow
enough that any lens with a separation greater than $\sim0\farcs2$
will be obvious by visual inspection, as we show explicitly in
Paper~I.  These observations should therefore be sensitive to
essentially all expected lenses.

\begin{figure}[p]
\epsscale{0.9}
\plotone{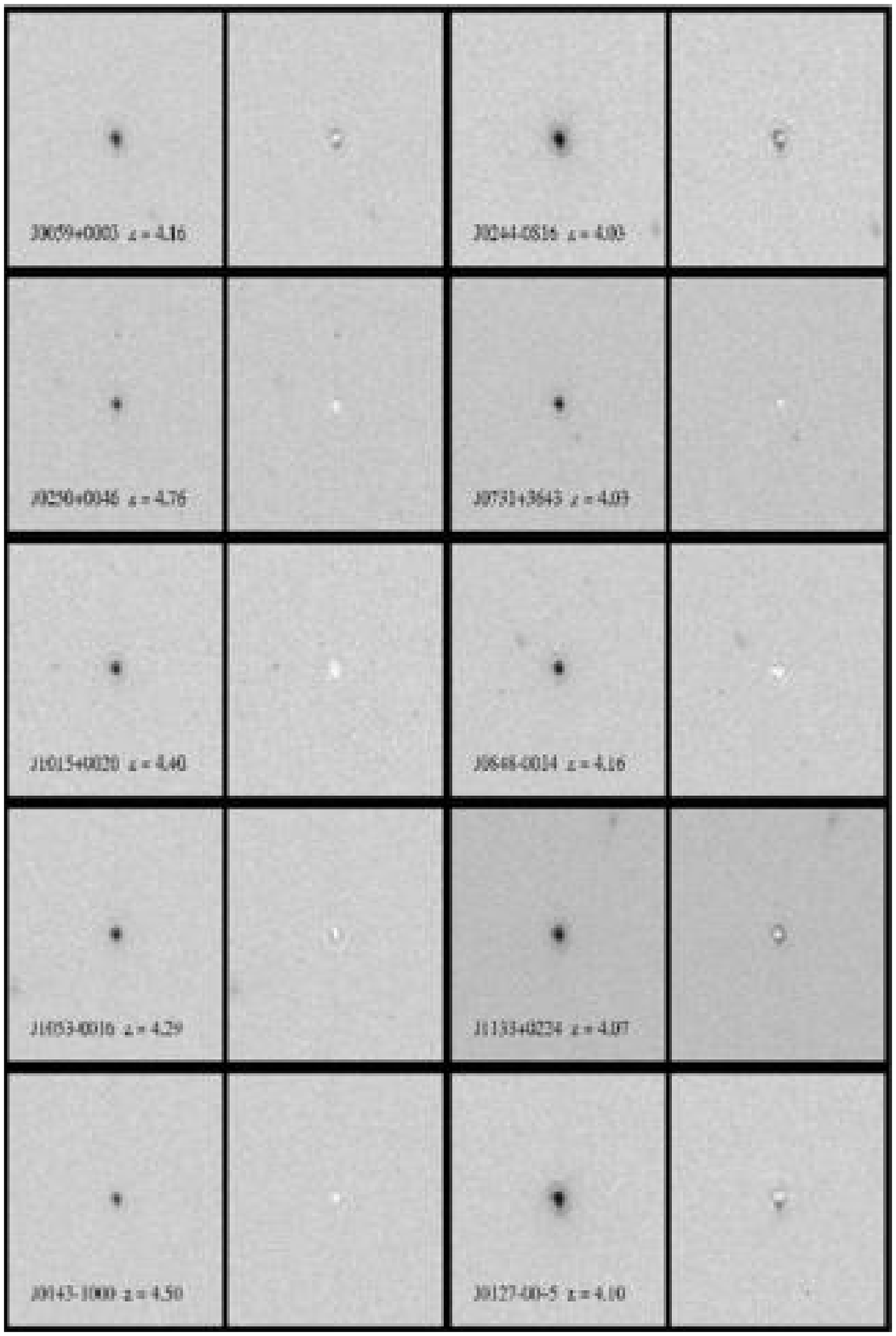}
\caption{{\em HST}/ACS/HRC F775W images of the 10 high-$z$ SDSS
quasars that show any sign of a secondary object in a $5\arcsec$ field
of view.  The scale is $5\arcsec\times5\arcsec$ in each of the
panels. The left hand panel is the ``cosmic ray rejected'' (CRJ)
output of CALACS.  The right hand panel is the same CRJ image after
subtraction of the Tiny Tim PSF (v6.1) with the same stretch as the
left hand panel.
\label{fig:fig1}}
\end{figure}

We search for faint secondary images at separations smaller than a few
tenths of an arcsecond by fitting and removing a model for the point
spread function of each image.  We use version v6.1 (which includes
on-orbit updates for ACS) of the Tiny Tim software
\markcite{kri95}({Krist} 1995)\footnote{http://www.stsci.edu/software/tinytim/}, which
produces a model PSF for the instruments on {\em HST} given the
object's (observed) spectral energy distribution and position in the
focal plane, the filter curve, and knowledge of the optics of the
instrument.  We fit the PSF model to each CRJ image, allowing the
location on the CCD and normalization to vary, and using sinc
interpolation when shifting the model PSF by fractional pixels.
Example PSF-subtracted images are shown in Figure~\ref{fig:fig1}.
Each of the images shows the familiar first Airy ring; on a very hard
stretch, the second Airy ring is faintly visible.

As discussed in Paper I, for a secondary image offset by $0\farcs1$
with a flux ratio of 10:1, the secondary object is visible as an
enhancement in the first Airy ring of the primary object; it becomes
clear upon PSF subtraction (Fig.~\ref{fig:fig1}).  With a flux ratio
of 100:1, we cannot discern a pair with $0\farcs1$ separation even
after subtraction of the PSF; however, even if we were able to do so,
it would not significantly improve the constraints that we derive
below since most split images will have larger image separations.

None of our 157 targets appears to be gravitationally lensed.
Figure~\ref{fig:fig1} shows only the ten objects (both with and
without PSF subtraction) which displayed any hint of a second object
within the 5\arcsec\ field of view.  Based on their brightness and
morphology, these secondary objects are most likely to be
galaxies along the line of sight to the target objects or (in some
cases) uncorrected cosmetic defects.

\subsection{Accounting for Extended-Source Selection Effects}

Since a gravitationally lensed pair of quasars with separation roughly
comparable to the seeing size may be classified as extended (rather
than point) sources, we must account for the effect that a
morphological selection criterion has on our analysis.  For the very
highest redshift ($z\gtrsim5$) quasars found from SDSS imaging
\markcite{fsr+01}(e.g., {Fan} {et~al.} 2001a, and references therein),
there is no selection bias against lensed quasars since no morphology
restriction was imposed.  However, for somewhat lower redshift
sources, which were selected with the automated pipeline
\markcite{rfn+02}({Richards} {et~al.} 2002), there is a strong bias
against high-$z$ ($z\gtrsim3$) quasars that appear extended (whether
due to lensing or errors in morphological classification as a result
of low S/N).  This bias is intentional; the survey cannot afford the
fibers that would be needed to explore all of the extended sources in
the high-$z$ quasar portion of the SDSS color-color diagram.  The vast
majority of these objects are faint moderate-redshift ($0.4<z<1$)
galaxies.

As discussed in Paper I and \markcite{ptl+03}{Pindor} {et~al.} (2003),
in order to test the ability of the SDSS photometric pipeline to
identify quasar pairs, we created simulated SDSS images of pairs of
point sources in these observing conditions and at the appropriate
signal-to-noise ratio.  The SDSS star-galaxy separator would have
classified as a galaxy, and hence excluded from the spectroscopic
sample, any pair of point sources having both an image separation
$\sim 1 \arcsec < \Delta \theta < \sim 2 \arcsec$, and a flux ratio
less than $\sim$ 5:1.  Thus the SDSS data are suitable for exploring
only a fraction of the parameter space of lensing (in terms of
separations and flux ratios) that is of interest.

The morphological selection likelihood depends on the luminosity and
on the assumed lensing flux ratio and splitting angle, for any given
individual quasar.  We therefore fold this bias into our probability
calculations below on an object--by--object basis.  For a given quasar
in our sample that has an apparent magnitude of $i_s$ at redshift
$z_s$, and for a given hypothesized magnification factor $\mu_s$
(which, for a singular isothermal sphere [SIS], implies a flux ratio
of $(\mu_s+2)/(\mu_s-2)$) for a lensing halo of mass $M$ (which
determines the splitting angle $\theta_s$), we compute the probability
$P_s(i_s,z_s,\mu_s,M)$ that the object would have been classified as
extended and discarded by the photometric pipeline.

To compute the impact of the selection on our final constraints, we
compute the product of the selection likelihood with the expected
lensing probability, which gives the probability that this quasar
would have been selected by SDSS {\it and} strongly lensed: $P_{\rm
lens}(z,M)=[1-P_s(i_s,z_s,\mu_s,M)] \times P_{\rm lens,0}(\mu_s,M)$,
where $P_{\rm lens,0}$ is the probability of multiply-imaged lensing
in the absence of any other selection effects.  Finally, we integrate
this product over all combinations of magnifications and lensing halo
masses that yield two detectable images with {\it HST}.  We find that
typically the inclusion of the morphological selection reduces the
total lensing probability by a factor of three (i.e., the value of the
above--defined integral would be three times larger if the factor
$P_s(i_s,z_s,\mu_s,M)$ was excluded).  More specifically, we find that
only $\sim$ 1/3 of true lenses would have made it into our sample,
with $\sim$ 2/3 being classified as extended sources.  In Table 1, we
list the overall lensing probability for each source.  Sixteen of the
157 sources were selected without regard to morphology (as indicated
in Table~1); we therefore did not apply the selection function to
these objects.

We note that there is a further possible selection effect due to the
fact that PSF magnitudes do not contain 100\% of the flux from a
gravitationally lensed pair of images, hence effectively reducing the
magnification bias. This effect turns out to be less significant,
partly because those lensed objects for which it would be most
important are already removed from the sample by the star-galaxy
morphological bias.  Overall, we find the effect is likely to shift
the luminosity scale by $\sim $0.1 magnitudes or less, and we ignore
it in what follows.

\section{Constraining the Slope of the Quasar Luminosity Function}

As discussed in Paper I, existing constraints on the high-redshift QLF
are limited by sample size and to the most luminous objects.
\markcite{ssg95}{Schmidt} {et~al.} (1995), using a set of 90 quasars with $2.7 < z < 4.8$, found a
power-law luminosity function slope ($\beta$) of roughly $-2$.
\markcite{fss+01}{Fan} {et~al.} (2001b) measured $\beta \approx -2.5\pm0.3$ from 39 quasars in
the redshift interval 3.6 to 5.0.  However, this relatively shallow
apparent slope does not necessarily represent the intrinsic slope of
the QLF, which is expected to be much steeper if a substantial
fraction of these quasars are magnified by lensing
\markcite{sef92}(e.g., {Schneider}, {Ehlers}, \& {Falco} 1992).

In line with Paper~I and previous work, we describe the intrinsic (not
necessarily observed) quasar luminosity function as a broken power law
\markcite{bsp88,pei95a}(e.g., {Boyle}, {Shanks}, \& {Peterson} 1988; {Pei} 1995):
\begin{equation}
\Phi_{\rm int}(L)=\frac{\Phi_\ast/L_\ast}{
(L/L_\ast)^{-\beta_l}+(L/L_\ast)^{-\beta_h}}.
\label{eq:qlf}
\end{equation}
The QLF is described by four parameters: the normalization
$\Phi_\ast$, the faint-end slope $\beta_l$, the bright-end slope
$\beta_h$, and the characteristic luminosity $L_\ast$ at which the QLF
steepens. The lensing probability is most sensitive to the last two
parameters, $\beta_h$ and $L_\ast$.  The faint end slope has
negligible impact on our analsysis and we set it to $\beta_l=-1.64$
\markcite{pei95a}(e.g., {Pei} 1995).  We apply the lensing model from
\markcite{chs02}{Comerford} {et~al.} (2002), in which lenses are
associated with dark matter halos, to compute the total lensing
probability, including the effect of magnification bias.  In this
model, the abundance of halos as a function of potential well depth is
adopted from the simulations of \markcite{jfw+01}{Jenkins} {et~al.}
(2001).  As discussed in Paper~I, we assume that all halos below
$M\approx 10^{13}~{\rm M_\odot}$ have SIS profiles (adopting a
standard conversion between circular velocity and halo mass), while
all halos above this mass follow the dark matter density profile
suggested by \markcite{nfw97}{Navarro}, {Frenk}, \& {White} (1997,
hereafter NFW).  NFW profiles are much less efficient lenses than are
SIS profiles.  This prescription is essentially equivalent to ignoring
all lenses above a halo mass of $10^{13} {\rm M_\odot}$, as the
massive halos do not contribute to lensing at small separations.  We
do not include here more complex lens models, such as those including
external shear or ellipticity.  As discussed by
\markcite{kkh05}{Keeton}, {Kuhlen}, \& {Haiman} (2005), such models
can, in general, boost the lensing probabilities (although that paper
considered in detail only the boost that can occur for singly--imaged
lensing probabilities).

In the case of $\beta_h=-3.8$ and $M^*_{1450}=-24.52$, the mean
probability for lensing is about 4\% for each source.  The morphology
selection effect causes this mean probability to be reduced by about a
factor of 3 (as discussed above).  In Table~\ref{tab:tab1} we list
each of the quasars that were observed by {\em HST} and list their
lensing probabilities, which vary relatively little from source to
source.  The median redshift of the sample is $z=4.35$.

\begin{figure}[p]
\epsscale{1.0}
\plotone{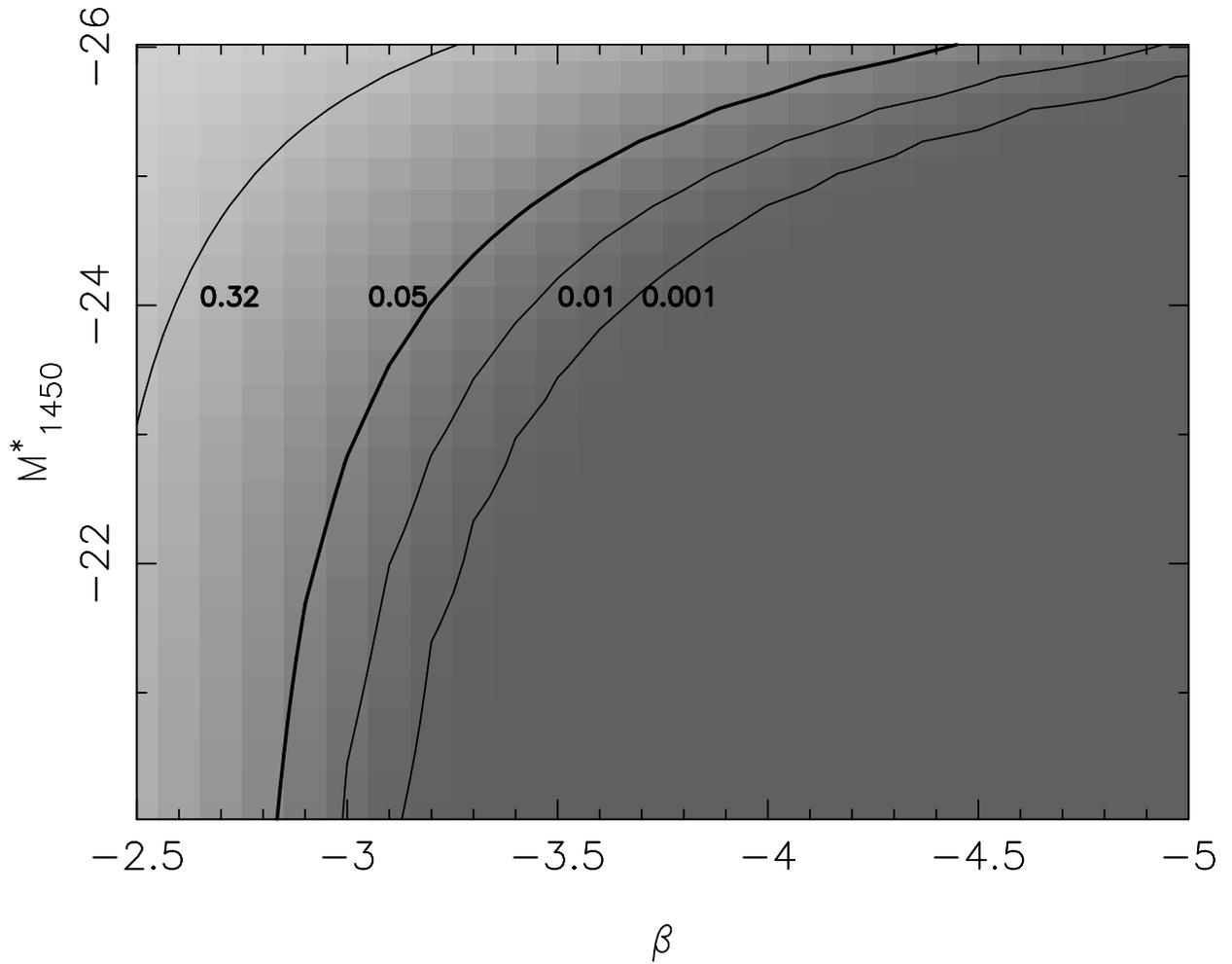}
\caption{Contours of fixed likelihood for no lensing among the 157
$z\sim4.35$ quasars, shown in the two--dimensional parameter space of
the slope and break of the $z\sim 4.35$ quasar luminosity function.
The $3\sigma$ limit on $\beta$ given an assumed break in the
luminosity function of $M_{1450}^\ast=-24.5$ is $\beta=-3.8$.
\label{fig:fig2}}
\end{figure}

The lack of lenses among these 157 quasars allows us to place
constraints on the QLF at high redshift, shown in
Figure~\ref{fig:fig2}.  The lensing probability is a function of both
the break and the faint-end slope.  Assuming a break of
$M_B^\ast=-25.0$ ($M_{1450}^\ast=-24.52$), the above lensing
probabilities yield a 3--$\sigma$ constraint on the bright end slope
of $\beta>-3.8$.\footnote{This constraint weakens if the intrinsic
bright end slope steepens with luminosity instead of being a single
power law.}  Note that we have used a somewhat brighter break
luminosity than we did in Paper I for the $z\sim6$ quasars as the
characteristic luminosity is thought to evolve with redshift.  Since
this choice of break luminosity is somewhat arbitrary (as it is an
extrapolation from low-redshift and the QLF may not even exhibit a
strong break at high-$z$), we also show the constraints on $\beta$ as
a function of break luminosity in Figure~\ref{fig:fig2}.

Finally, it is useful to consider constraints that we can place on the
underlying lensing halo population. Without any magnification bias,
the total optical depth to multiply-imaged lensing at $z\sim4.5$ is
only 0.25 percent, which is reduced by the SDSS morphological
selection to 0.08 percent.  This makes the lack of lenses among our
157 sources unsurprising.  We find that to reduce the probability of
not finding even one lens in the whole sample, the single-object
lensing probability must be boosted by a factor of about 40 (this is
essentially the typical magnification bias produced by the model QLF
we constrain at 3$\sigma$).  As a result, we rule out (at 3$\sigma$
confidence) lensing models in which there are $\gsim 40$ times more
galaxies than implied by the halo mass function we adopted from
Jenkins et al.  While this is a weak constraint, it is still the best
direct limit on the number density of $M\sim 10^{12}~{\rm M_\odot}$
halos at $1<z<2$ (the redshift and mass range dominating the expected
lensing probability; see Figure 1 in \markcite{chs02}{Comerford}
{et~al.} 2002).

\section{Discussion and Conclusions}

Constraining the high-$z$ QLF slope is particularly important to
understand the roles of accretion and feedback in the growth of
galaxies \markcite{sr98,fab99,wl03b,hhc+05a}(e.g., {Silk} \& {Rees}
1998; {Fabian} 1999; {Wyithe} \& {Loeb} 2003b; {Hopkins et al.}
2005a).  For the most luminous quasars with $z\lesssim2.5$ (i.e., the
redshift at which the quasar comoving density peaks), the bright end
slope has been shown to be $\beta\sim-3.3$
\markcite{csb+04,rca+05}({Croom} {et~al.} 2004; {Richards} {et~al.}
2005).  For the most luminous quasars at $z>4$, the measured slope is
$\beta\sim-2.5$ \markcite{fsr+01}({Fan} {et~al.} 2001a).  While the
slope of the QLF has been well measured, understanding the physical
processes behind the QLF and its evolution are still open questions
and various explanations have been proposed.  \markcite{wl03}{Wyithe}
\& {Loeb} (2003a) suggest that feedback mechanisms may prevent the gas
in the most massive dark matter halos from collapsing at low-$z$,
suppressing the number of luminous low-$z$ quasars relative to
high-$z$.  Alternatively, in the model of
\markcite{hhc+05a,hhc+05b}{Hopkins et al.} (2005a, 2005b), the bright
end of the QLF is defined by near-Eddington accretion at the peak
luminosity in the history of a quasar, whereas the faint end slope may
be due to sub-Eddington accretion.  Tighter limits on the bright end
slope of the high-redshift QLF provide important constraints that any
such models must account for.

In this work, we have obtained high resolution {\em HST} images of 157
$4<z<5.4$ redshift quasars (known prior to September 2001) to look for
the signature of gravitational lensing.  We have found no evidence of
multiple images, significantly limiting the amount by which these
quasars can be magnified by foreground mass concentrations (in the
absence of microlensing).  The lack of any strong lenses puts a
$3\sigma$ constraint on the intrinsic bright end slope of the $z=$4--5
luminosity function of $\beta_h>-3.8$.  Our sample has a strong bias
against pairs with $\theta>1\arcsec$ due to the exclusion of objects
which appear extended in SDSS images.  We are currently exploring
methods for cleanly selecting such wide separation pairs from the SDSS
imaging data.

\acknowledgements Funding for the creation and distribution of the
SDSS Archive has been provided by the Alfred P. Sloan Foundation, the
Participating Institutions, the National Aeronautics and Space
Administration, the National Science Foundation, the U.S. Department
of Energy, the Japanese Monbukagakusho, and the Max Planck
Society. The SDSS Web site is http://www.sdss.org/.  The SDSS is
managed by the Astrophysical Research Consortium (ARC) for the
Participating Institutions. The Participating Institutions are The
University of Chicago, Fermilab, the Institute for Advanced Study, the
Japan Participation Group, The Johns Hopkins University, the Korean
Scientist Group, Los Alamos National Laboratory, the
Max-Planck-Institute for Astronomy (MPIA), the Max-Planck-Institute
for Astrophysics (MPA), New Mexico State University, University of
Pittsburgh, University of Portsmouth, Princeton University, the United
States Naval Observatory, and the University of Washington.  Support
for program \#9472 was provided by NASA through a grant
(HST-GO-09472.01-A; M.~A.~S and G.~T.~R.) from the Space Telescope
Science Institute, which is operated by the Association of
Universities for Research in Astronomy, Inc., under NASA contract NAS
5-26555.  This work was supported in part by NASA ATP grant NNG04GI88G
(Z.~H.), National Science Foundation grants AST-0307582 (D.~P.~S.),
AST-0307409 (M.~A.~S.), AST-0307384 (X.~F.), AST-0307291 (Z.~H.) and
AST-0307200 (Z.~H.).  X.~F. and D.~J.~E. acknowledge Alfred P. Sloan
Research Fellowships.  X.~F. further acknowledges support from a David
and Lucile Packard Fellow in Science and Engineering.




\clearpage

\begin{deluxetable}{lllllllll}
\tabletypesize{\scriptsize}
\tablewidth{0pt}
\tablecaption{\label{tab:tab1}}
\tablehead{
\colhead{SDSS~J} &
\colhead{Redshift} &
\colhead{$M_B$} &
\colhead{SDSS $i$} &
\colhead{HST $i$} &
\colhead{$P_{\rm lens,0}$} &
\colhead{$P_{\rm lens}$} &
\colhead{HST ID} &
\colhead{Ref}
}
\startdata
001115.23+144601.8 & 4.924 & $-28.39$ & 18.41 & 18.26 & 0.3137 & 0.0906 & aa & 7\\
001134.52+155137.3 & 4.394 & $-26.20$ & 20.20 & 19.91 & 0.0553 & 0.0164 & ab & 7\\
001714.66$-$100055.4 & 4.976 & $-26.85$ & 19.61 & 19.46 & 0.1103 & 0.0277 & ac & 7\\
001813.88+142455.6 & 4.221 & $-26.81$ & 19.41 & 19.30 & 0.0945 & 0.0269 & ad & 7\\
001918.43+150611.3 & 4.134 & $-26.27$ & 19.91 & 19.94 & 0.0563 & 0.0160 & ae & 7\\
001950.06$-$004040.7 & 4.352 & $-26.65$ & 19.50 & 19.36 & 0.0842 & 0.0345 & af & 3\\
002618.67+140946.7 & 4.578 & $-26.64$ & 20.11 & 19.68 & 0.0868 & 0.0221 & ag & 7\\
003525.28+004002.7$^*$ & 4.750 & $-26.77$ & 19.42 & 19.90 & 0.0999 & 0.0999 & ai & 1\\
003618.84$-$004629.9 & 4.040 & $-26.46$ & 19.89 & 19.57 & 0.0662 & 0.0175 & aj & 0\\
003714.11$-$005603.9$^*$ & 4.350 & $-26.49$ & 20.27 & 20.55 & 0.0727 & 0.0727 & ak & 6\\
003749.19+155208.4 & 4.076 & $-26.05$ & 19.72 & 19.72 & 0.0443 & 0.0182 & al & 6\\
004054.65$-$091526.8 & 4.972 & $-27.73$ & 19.35 & 19.05 & 0.2124 & 0.0531 & am & 7\\
005006.35$-$005319.2 & 4.416 & $-26.95$ & 19.53 & 19.39 & 0.1104 & 0.0303 & an & 6\\
005922.65+000301.4 & 4.161 & $-26.97$ & 19.43 & 19.24 & 0.1069 & 0.0394 & ao & 3\\
010619.25+004823.4 & 4.430 & $-27.73$ & 18.70 & 18.59 & 0.1980 & 0.0546 & aq & 6\\
012004.82+141108.2 & 4.720 & $-25.78$ & 20.22 & 20.45 & 0.0375 & 0.0120 & ar & 6\\
012019.99+000735.5 & 4.085 & $-26.31$ & 19.82 & 19.87 & 0.0579 & 0.0230 & as & 3\\
012211.11+150914.3 & 4.510 & $-27.60$ & 18.67 & 18.65 & 0.1835 & 0.0574 & at & 7\\
012405.70$-$004407.8 & 4.008 & $-26.57$ & 19.56 & 19.42 & 0.0728 & 0.0200 & au & 7\\
012700.69$-$004559.2 & 4.103 & $-27.91$ & 18.30 & 18.11 & 0.2105 & 0.0668 & av & 3\\
013242.76$-$094301.6 & 4.268 & $-26.62$ & 19.71 & 19.53 & 0.0807 & 0.0211 & aw & 7\\
014328.37$-$100019.3 & 4.496 & $-26.36$ & 20.10 & 20.16 & 0.0660 & 0.0190 & ax & 7\\
014609.33$-$092918.2 & 4.153 & $-26.32$ & 19.92 & 19.73 & 0.0593 & 0.0168 & ay & 7\\
015339.60$-$001104.8 & 4.205 & $-27.23$ & 18.74 & 18.81 & 0.1329 & 0.0399 & a0 & 6\\
015704.10+122858.2 & 4.191 & $-26.06$ & 19.90 & 19.90 & 0.0460 & 0.0151 & a1 & 6\\
020326.46$-$003954.0 & 4.169 & $-27.02$ & 19.45 & 19.15 & 0.1116 & 0.0315 & a3 & 0\\
020651.37+121624.3$^*$ & 4.810 & $-26.79$ & 19.89 & 19.84 & 0.1025 & 0.1025 & a4 & 6\\
021043.15$-$001818.2 & 4.697 & $-27.11$ & 19.54 & 19.14 & 0.1313 & 0.0406 & a5 & 3\\
021102.72$-$000910.2$^*$ & 4.900 & $-27.05$ & 19.93 & 20.04 & 0.1286 & 0.1286 & a6 & 1\\
023137.65$-$072854.5 & 5.410 & $-69.14$ & 19.53 & 19.55 & 0.0073 & 0.0029 & a8 & 6\\
023923.47$-$081005.1 & 4.019 & $-26.46$ & 19.27 & 19.18 & 0.0658 & 0.0204 & a9 & 6\\
024447.78$-$081606.1 & 4.030 & $-28.03$ & 18.03 & 18.19 & 0.2240 & 0.0785 & ba & 6\\
024457.19$-$010809.8 & 4.010 & $-26.46$ & 18.33 & 18.26 & 0.0657 & 0.0266 & bb & 3\\
025019.78+004650.3$^*$ & 4.760 & $-27.03$ & 19.65 & 19.72 & 0.1243 & 0.1243 & bc & 3\\
025647.06$-$085041.3 & 4.228 & $-26.95$ & 19.62 & 19.65 & 0.1066 & 0.0308 & bf & 6\\
030025.22+003224.3 & 4.188 & $-26.12$ & 20.18 & 19.90 & 0.0489 & 0.0192 & bg & 3\\
031213.97$-$062658.8 & 4.030 & $-27.12$ & 19.08 & 19.08 & 0.1174 & 0.0341 & bj & 6\\
032608.12$-$003340.1 & 4.175 & $-26.89$ & 19.26 & 19.51 & 0.1003 & 0.0279 & bl & 1\\
033119.66$-$074143.1 & 4.700 & $-27.21$ & 19.10 & 19.35 & 0.1420 & 0.0393 & bm & 6\\
033305.32$-$053708.9 & 4.270 & $-26.07$ & 19.79 & 19.83 & 0.0473 & 0.0172 & bn & 6\\
033829.31+002156.2 & 5.001 & $-26.29$ & 20.15 & 19.96 & 0.0668 & 0.0199 & bo & 1\\
035214.33$-$001941.0$^*$ & 4.180 & $-26.46$ & 19.76 & 19.78 & 0.0683 & 0.0683 & bs & 3\\
040550.26+005931.1$^*$ & 4.050 & $-25.85$ & 20.07 & 20.07 & 0.0356 & 0.0356 & bt & 3\\
043225.11$-$005625.9$^*$ & 4.020 & $-25.47$ & 19.77 & 19.89 & 0.0230 & 0.0230 & bu & 0\\
073147.00+364346.5 & 4.030 & $-27.12$ & 19.29 & 19.22 & 0.1174 & 0.0369 & bv & 6\\
074640.16+344624.7 & 4.032 & $-26.73$ & 19.34 & 19.31 & 0.0846 & 0.0259 & bx & 7\\
075103.95+424211.5 & 4.160 & $-27.35$ & 18.82 & 18.87 & 0.1443 & 0.0450 & bz & 6\\
075618.14+410408.5 & 5.090 & $-26.60$ & 20.21 & 20.07 & 0.0915 & 0.0234 & b0 & 6\\
075652.07+450258.9 & 4.812 & $-26.37$ & 20.24 & 20.23 & 0.0700 & 0.0199 & b1 & 6\\
075732.90+441424.6 & 4.179 & $-26.94$ & 19.38 & 19.42 & 0.1047 & 0.0300 & b2 & 6\\
081054.88+460357.8 & 4.087 & $-27.46$ & 18.38 & 18.53 & 0.1542 & 0.0478 & b5 & 6\\
083100.68+434426.9 & 4.452 & $-26.42$ & 20.08 & 19.95 & 0.0694 & 0.0188 & b7 & 7\\
083103.00+523533.5 & 4.512 & $-27.00$ & 19.37 & 19.28 & 0.1169 & 0.0329 & b8 & 6\\
083212.38+530327.4 & 4.081 & $-26.74$ & 19.46 & 19.28 & 0.0863 & 0.0258 & b9 & 6\\
083941.45+031817.0 & 4.244 & $-27.61$ & 18.79 & 18.55 & 0.1769 & 0.0543 & cc & 7\\
083946.21+511202.8 & 4.415 & $-27.71$ & 18.98 & 18.83 & 0.1949 & 0.0597 & cd & 6\\
084811.52$-$001417.9 & 4.156 & $-27.38$ & 18.99 & 18.93 & 0.1474 & 0.0456 & ce & 6\\
085151.26+020756.0 & 4.294 & $-26.75$ & 19.41 & 19.27 & 0.0910 & 0.0250 & cf & 6\\
085210.89+535948.9 & 4.275 & $-26.45$ & 19.82 & 19.77 & 0.0690 & 0.0189 & cg & 6\\
085227.28+504510.8 & 4.219 & $-27.70$ & 18.82 & 18.63 & 0.1873 & 0.0546 & ch & 7\\
090100.61+472536.1 & 4.610 & $-26.70$ & 19.76 & 19.56 & 0.0920 & 0.0275 & ck & 7\\
090440.63+535038.8 & 4.298 & $-27.27$ & 19.25 & 19.48 & 0.1394 & 0.0393 & cm & 6\\
090634.84+023433.8 & 4.544 & $-27.71$ & 18.77 & 18.66 & 0.1987 & 0.0712 & co & 7\\
091316.55+591921.4 & 5.110 & $-26.38$ & 20.50 & 20.47 & 0.0751 & 0.0179 & cq & 6\\
092038.49+564235.9 & 4.183 & $-26.62$ & 19.95 & 19.06 & 0.0793 & 0.0239 & cr & 6\\
092256.19+561849.1 & 4.198 & $-27.19$ & 18.99 & 19.67 & 0.1286 & 0.0403 & cs & 6\\
093931.90+003955.0$^*$ & 4.500 & $-26.74$ & 20.62 & 20.29 & 0.0936 & 0.0936 & cv & 5\\
094056.02+584830.1 & 4.660 & $-27.06$ & 19.27 & 19.41 & 0.1255 & 0.0361 & cw & 6\\
094108.35+594725.8 & 4.820 & $-27.10$ & 19.36 & 19.25 & 0.1325 & 0.0369 & cx & 6\\
094917.18+602104.4 & 4.290 & $-26.01$ & 20.26 & 20.20 & 0.0446 & 0.0136 & cy & 6\\
095151.17+594556.2 & 4.860 & $-26.41$ & 19.81 & 19.73 & 0.0732 & 0.0218 & c0 & 6\\
095511.33+594030.7 & 4.356 & $-27.68$ & 18.30 & 18.56 & 0.1892 & 0.0552 & c1 & 7\\
101053.52+644832.1 & 4.745 & $-26.35$ & 19.87 & 19.98 & 0.0680 & 0.0182 & c3 & 6\\
101549.00+002019.9 & 4.399 & $-27.02$ & 19.27 & 19.23 & 0.1166 & 0.0336 & c4 & 6\\
102043.82+000105.7 & 4.160 & $-26.08$ & 19.83 & 19.88 & 0.0466 & 0.0182 & c5 & 6\\
102119.16$-$030937.1$^*$ & 4.696 & $-25.88$ & 20.08 & 20.29 & 0.0417 & 0.0417 & c6 & 4\\
102332.08+633508.1 & 4.880 & $-27.00$ & 19.69 & 19.44 & 0.1232 & 0.0342 & c7 & 6\\
103432.72$-$002702.5 & 4.380 & $-26.17$ & 19.90 & 20.05 & 0.0535 & 0.0150 & c9 & 7\\
104008.10+651429.2 & 4.583 & $-26.22$ & 19.83 & 19.83 & 0.0584 & 0.0183 & da & 6\\
104040.14$-$001540.8 & 4.320 & $-27.40$ & 18.80 & 18.94 & 0.1543 & 0.0532 & db & 6\\
104351.19+650647.6 & 4.542 & $-27.15$ & 19.09 & 18.89 & 0.1324 & 0.0371 & dc & 6\\
105320.43$-$001649.5 & 4.291 & $-27.01$ & 19.33 & 19.30 & 0.1134 & 0.0343 & dg & 2\\
105902.73+010404.1 & 4.060 & $-26.89$ & 19.21 & 19.32 & 0.0978 & 0.0292 & di & 6\\
110247.29+663519.5 & 4.810 & $-26.12$ & 20.47 & 20.57 & 0.0546 & 0.0142 & dj & 6\\
110813.86$-$005944.5 & 4.029 & $-26.73$ & 19.25 & 19.41 & 0.0845 & 0.0243 & dk & 5\\
110826.31+003706.7 & 4.502 & $-26.79$ & 19.84 & 19.55 & 0.0978 & 0.0348 & dm & 6\\
111224.18+004630.3 & 4.032 & $-26.34$ & 19.66 & 19.62 & 0.0588 & 0.0172 & dn & 6\\
111401.47$-$005321.0 & 4.590 & $-26.70$ & 19.57 & 19.70 & 0.0917 & 0.0271 & do & 2\\
112242.99$-$022905.1$^*$ & 4.795 & $-26.40$ & 20.38 & 20.43 & 0.0719 & 0.0719 & dp & 4\\
112253.51+005329.7 & 4.570 & $-26.80$ & 19.12 & 19.28 & 0.0998 & 0.0409 & dq & 2\\
112311.13$-$004418.5$^*$ & 5.000 & $-26.28$ & 20.46 & 20.20 & 0.0661 & 0.0661 & dr & 0\\
113354.89+022420.9 & 4.066 & $-27.41$ & 18.84 & 18.88 & 0.1480 & 0.0444 & dt & 7\\
113559.93+002422.7 & 4.040 & $-26.31$ & 19.87 & 19.64 & 0.0572 & 0.0163 & du & 5\\
115547.83+022716.1 & 4.355 & $-26.81$ & 19.51 & 19.29 & 0.0970 & 0.0273 & dx & 7\\
120439.42+663549.7 & 4.052 & $-26.72$ & 19.66 & 19.62 & 0.0842 & 0.0252 & dy & 7\\
120441.73$-$002149.6 & 5.030 & $-27.19$ & 19.28 & 19.28 & 0.1471 & 0.0646 & dz & 2\\
120823.81+001027.6$^*$ & 5.280 & $-26.82$ & 20.79 & 20.59 & 0.1164 & 0.1164 & d1 & 1\\
121422.02+665707.5 & 4.680 & $-26.47$ & 18.91 & 18.96 & 0.0755 & 0.0238 & d2 & 7\\
122600.68+005923.5 & 4.264 & $-27.24$ & 18.90 & 18.83 & 0.1354 & 0.0524 & d3 & 2\\
122622.03+662017.9 & 4.017 & $-26.24$ & 20.07 & 20.05 & 0.0531 & 0.0192 & d4 & 7\\
122657.97+000938.4 & 4.140 & $-27.03$ & 19.24 & 19.31 & 0.1118 & 0.0356 & d5 & 7\\
123503.02$-$000331.6 & 4.700 & $-25.98$ & 20.07 & 20.04 & 0.0465 & 0.0134 & d8 & 2\\
125759.21$-$011130.2 & 4.150 & $-27.79$ & 18.55 & 18.40 & 0.1964 & 0.0613 & ec & 6\\
125847.62$-$025456.1 & 4.031 & $-26.68$ & 19.57 & 19.45 & 0.0000 & 0.0000 & ee & 7\\
130002.16+011823.0 & 4.614 & $-27.25$ & 18.82 & 18.73 & 0.1446 & 0.1446 & ef & 7\\
130619.38+023658.9 & 4.874 & $-26.92$ & 19.66 & 19.61 & 0.1153 & 0.0298 & ej & 7\\
131052.51$-$005533.3 & 4.150 & $-27.32$ & 18.83 & 18.85 & 0.1408 & 0.0451 & ek & 2\\
131831.83+653929.4 & 4.286 & $-26.09$ & 20.07 & 20.17 & 0.0484 & 0.0180 & el & 7\\
132110.82+003821.6 & 4.700 & $-26.46$ & 20.04 & 20.22 & 0.0750 & 0.0255 & em & 2\\
134134.19+014157.7 & 4.725 & $-27.51$ & 18.93 & 18.98 & 0.1776 & 0.0513 & eq & 7\\
134723.08+002158.8 & 4.270 & $-27.54$ & 18.96 & 18.71 & 0.1692 & 0.0499 & er & 6\\
140146.52+024434.6 & 4.455 & $-27.72$ & 18.58 & 18.53 & 0.1974 & 0.0574 & ev & 7\\
140248.07+014634.1 & 4.206 & $-27.96$ & 18.26 & 18.12 & 0.2215 & 0.0612 & ew & 7\\
141306.09+644149.0 & 4.255 & $-26.28$ & 19.49 & 19.34 & 0.0583 & 0.0170 & ey & 7\\
141315.36+000032.4 & 4.071 & $-26.34$ & 19.73 & 19.54 & 0.0594 & 0.0191 & ez & 2\\
142911.84+632344.9 & 4.431 & $-26.57$ & 20.33 & 20.33 & 0.0795 & 0.0182 & e4 & 7\\
143352.21+022714.0 & 4.749 & $-28.03$ & 18.33 & 18.32 & 0.2502 & 0.0763 & e5 & 7\\
144255.56+590949.9 & 4.333 & $-26.23$ & 20.19 & 20.04 & 0.0564 & 0.0198 & e8 & 7\\
144340.71+585653.3 & 4.242 & $-28.44$ & 18.02 & 18.09 & 0.2966 & 0.0864 & e9 & 7\\
144413.26+004836.7$^*$ & 4.780 & $-26.08$ & 21.00 & 21.08 & 0.0522 & 0.0522 & fb & 0\\
144617.35$-$010131.1 & 4.185 & $-27.21$ & 19.09 & 19.06 & 0.1303 & 0.0376 & fd & 6\\
144717.97+040112.4 & 4.508 & $-27.48$ & 19.33 & 19.21 & 0.1686 & 0.0449 & fe & 7\\
145107.94+025615.6 & 4.486 & $-27.05$ & 19.11 & 19.08 & 0.1212 & 0.0354 & ff & 7\\
145229.37+595156.3 & 4.029 & $-26.85$ & 19.54 & 19.32 & 0.0938 & 0.0317 & fi & 7\\
145350.38+610109.0 & 4.134 & $-26.59$ & 19.50 & 19.49 & 0.0763 & 0.0235 & fj & 7\\
145747.66+575332.1 & 4.355 & $-26.47$ & 20.04 & 19.93 & 0.0714 & 0.0230 & fk & 7\\
150527.34+573632.0 & 4.385 & $-26.36$ & 19.98 & 19.75 & 0.0647 & 0.0197 & fl & 7\\
150847.60+571501.3 & 4.876 & $-26.53$ & 20.06 & 19.99 & 0.0821 & 0.0240 & fm & 7\\
151002.92+570243.3 & 4.310 & $-26.43$ & 20.06 & 19.95 & 0.0682 & 0.0201 & fn & 7\\
151155.98+040802.9 & 4.621 & $-26.63$ & 19.85 & 19.71 & 0.0866 & 0.0321 & fp & 7\\
161616.25+513336.9 & 4.528 & $-26.98$ & 19.58 & 19.54 & 0.1153 & 0.0329 & fz & 7\\
163257.07+441110.2 & 4.105 & $-27.35$ & 18.93 & 18.54 & 0.1427 & 0.0431 & f1 & 7\\
163950.51+434003.7 & 4.007 & $-28.37$ & 17.91 & 17.79 & 0.2738 & 0.0860 & f2 & 7\\
165354.62+405402.2 & 4.966 & $-27.23$ & 18.79 & 18.54 & 0.1491 & 0.0481 & f3 & 7\\
170804.89+602201.9 & 4.350 & $-26.54$ & 19.78 & 19.99 & 0.0762 & 0.0225 & f4 & 6\\
171014.51+592326.4 & 4.536 & $-26.75$ & 19.67 & 19.73 & 0.0950 & 0.0288 & f5 & 6\\
171224.92+560624.9 & 4.229 & $-26.23$ & 20.02 & 20.07 & 0.0552 & 0.0195 & f6 & 6\\
171808.67+551511.2 & 4.620 & $-26.35$ & 19.99 & 20.06 & 0.0667 & 0.0201 & f7 & 6\\
172007.20+602823.8 & 4.400 & $-26.20$ & 20.20 & 20.08 & 0.0554 & 0.0188 & f8 & 6\\
173744.87+582829.5 & 4.940 & $-27.35$ & 19.33 & 19.29 & 0.1626 & 0.0476 & f9 & 6\\
204421.51$-$052521.9 & 4.224 & $-26.88$ & 19.71 & 19.52 & 0.1004 & 0.0232 & ga & 7\\
210155.45$-$062711.8 & 4.341 & $-26.09$ & 20.20 & 19.95 & 0.0490 & 0.0130 & gb & 7\\
210216.52+104906.5 & 4.175 & $-27.34$ & 19.22 & 18.96 & 0.1436 & 0.0470 & gc & 7\\
220008.66+001744.8 & 4.791 & $-26.93$ & 19.30 & 19.19 & 0.1150 & 0.0324 & gg & 6\\
220307.39$-$004612.0 & 4.160 & $-27.11$ & 19.41 & 18.99 & 0.1198 & 0.0506 & gh & 7\\
221644.01+001348.2$^*$ & 4.990 & $-27.42$ & 20.30 & 20.31 & 0.1726 & 0.1726 & gj & 6\\
221855.11+134708.6 & 4.277 & $-26.60$ & 19.83 & 19.53 & 0.0793 & 0.0219 & gl & 7\\
222509.17$-$001406.9 & 4.861 & $-27.32$ & 19.31 & 19.07 & 0.1576 & 0.0602 & gn & 7\\
222845.14$-$075755.2 & 5.151 & $-27.33$ & 20.16 & 19.83 & 0.1684 & 0.0576 & gp & 7\\
224243.03$-$091543.9 & 4.223 & $-26.27$ & 19.95 & 19.88 & 0.0574 & 0.0156 & gr & 7\\
224630.86+131706.7 & 4.116 & $-26.72$ & 19.70 & 19.65 & 0.0854 & 0.0223 & gs & 7\\
225843.27$-$092710.6 & 4.041 & $-26.88$ & 19.35 & 18.95 & 0.0965 & 0.0277 & gw & 7\\
234003.50+140257.2 & 4.535 & $-27.36$ & 19.12 & 18.91 & 0.1552 & 0.0467 & g1 & 7\\
234025.97+135009.1 & 4.173 & $-25.93$ & 20.18 & 19.93 & 0.0399 & 0.0111 & g2 & 7\\
234750.31+134102.7 & 4.265 & $-26.05$ & 19.75 & 19.63 & 0.0462 & 0.0152 & g3 & 7\\
235152.80+160048.9 & 4.668 & $-26.31$ & 19.79 & 19.83 & 0.0647 & 0.0189 & g4 & 7\\
235344.26+143525.2 & 4.257 & $-26.58$ & 19.87 & 19.66 & 0.0776 & 0.0206 & g5 & 7\\
\enddata
\tablecomments{Column 1 gives the IAU name of the object (N.B. that
SDSS image reprocessing can change the coordinates [by $\sim0\farcs1$]
and thus the name slightly; matching should thus be done on
coordinates, not names).  Starred objects were selected without
reference to their morphological type (stellar vs.\ extended).  Column
2 is the redshift.  Column 3 is the absolute B magnitude, computed
with the parameters given in $\S~1$.  Columns 4 and 5 give the SDSS
and HST $i$-band magnitudes (uncorrected for Galactic reddening).
Column 6 indicates the lensing probability for this object, ignoring
morphological selection effects.  Column 7 gives the lensing
probability after accounting for morphological selection effects.
Column 8 give the two character ``observation set id'' related to the
{\em HST} file naming convention.  This is given to facilitate
archival use of these data.  The program id for this project was
``8f3'', thus the files for the first object have names like
j8f3aa011\_crj.fits.  Column 9 gives the discovery reference, which
are (1) \markcite{fss+99}{Fan} {et~al.} (1999), (2) \markcite{fss+00}{Fan} {et~al.} (2000), (3) \markcite{fsr+01}{Fan} {et~al.} (2001a), (4)
\markcite{zts+00}{Zheng} {et~al.} (2000), (5) \markcite{sfs+01}{Schneider} {et~al.} (2001), (6) \markcite{afr+01}{Anderson} {et~al.} (2001), and (7)
\markcite{shr+05}{Schneider et al.} (2005).  A zero in this column means that the objects are
being published here for the first time.}
\end{deluxetable}

\begin{deluxetable}{llll|llll}
\tabletypesize{\scriptsize}
\tablewidth{0pt}
\tablecaption{\label{tab:tab2}}
\tablehead{
\colhead{SDSS~J} &
\colhead{Redshift} &
\colhead{SDSS $i$} &
\colhead{Ref} &
\colhead{SDSS~J} &
\colhead{Redshift} &
\colhead{SDSS $i$} &
\colhead{Ref}
}
\startdata
003126.80+150739.6 & 4.291 & 19.97 & 7 &       132853.65$-$022441.6 & 4.620 & 19.91 & 7\\     
010326.89+005538.6 & 4.159 & 20.07 & 0 &       133211.90+031556.3 & 4.727 & 19.29 & 7\\       
015032.87+143425.5 & 4.284 & 20.09 & 6 &       135057.86$-$004355.3 & 4.427 & 19.91 & 6\\     
020152.53$-$094733.4 & 4.026 & 20.29 & 7 &     135134.46$-$003652.1 & 4.034 & 19.88 & 6\\     
021419.42$-$010716.9 & 4.592 & 20.53 & 0 &     135422.99$-$003906.1 & 4.420 & 20.15 & 6\\     
025039.17$-$065405.1 & 4.505 & 19.84 & 6 &     140404.63+031403.9 & 4.969 & 19.53 & 7\\       
025204.28+003136.9 & 4.119 & 19.91 & 6 &       141332.35$-$004909.6 & 4.213 & 19.30 & 2\\     
030437.21+004653.6 & 4.281 & 20.12 & 0 &       141534.91+033132.1 & 4.451 & 19.89 & 7\\       
031036.96$-$001457.0$^*$ & 4.630 & 19.89 & 1 & 142004.11+022708.7 & 4.187 & 20.09 & 7\\       
032459.10$-$005705.1$^*$ & 4.800 & 20.69 & 3 & 142408.34+024219.9 & 4.342 & 19.79 & 7\\       
034109.35$-$064805.0 & 4.142 & 20.20 & 6 &     144117.46+035910.5 & 4.312 & 19.60 & 7\\       
034541.51$-$072315.3 & 4.062 & 19.52 & 6 &     144231.72+011055.3 & 4.560 & 19.93 & 6\\       
034946.61$-$065730.2 & 4.041 & 20.23 & 6 &     144407.63$-$010152.7 & 4.552 & 19.29 & 6\\     
073354.93+321241.5 & 4.446 & 20.38 & 7 &       144428.67$-$012344.0 & 4.160 & 19.47 & 2\\     
074907.57+355543.8 & 4.257 & 20.09 & 7 &       145118.77$-$010446.1$^*$ & 4.660 & 20.70 & 1\\ 
080159.24+433624.9 & 4.166 & 20.26 & 6 &       145212.86+023526.4 & 4.916 & 19.92 & 7\\       
080549.94+482345.8 & 4.208 & 19.86 & 6 &       151041.79+031810.5 & 4.230 & 19.63 & 7\\       
081241.12+442129.0 & 4.334 & 20.21 & 6 &       151909.09+030633.7 & 4.391 & 19.47 & 7\\       
083824.32+460443.7 & 4.011 & 19.60 & 7 &       152245.19+024543.8 & 4.087 & 18.91 & 7\\       
085430.19+004213.6 & 4.079 & 20.02 & 6 &       152443.19+011358.9$^*$ & 4.114 & 19.98 & 6\\   
085634.93+525206.3 & 4.790 & 20.31 & 6 &       152740.50$-$010602.5 & 4.410 & 19.95 & 2\\     
090242.09$-$002125.8 & 4.450 & 19.77 & 6 &     152743.86+035301.3 & 4.226 & 19.86 & 7\\       
090532.14$-$001430.4 & 4.254 & 19.89 & 6 &     153259.95$-$003944.0$^*$ & 4.620 & 19.73 & 2\\ 
091016.79+575331.0 & 4.005 & 19.50 & 6 &       160207.95+523717.9 & 4.898 & 19.85 & 7\\       
092303.53+024739.4 & 4.669 & 20.39 & 7 &       160501.21$-$011220.0 & 4.920 & 19.78 & 2\\     
092819.28+534024.2 & 4.413 & 19.67 & 7 &       161544.13+010401.8$^*$ & 4.013 & 20.20 & 5\\   
095000.17+620318.5 & 4.062 & 20.22 & 6 &       162048.74+002005.7 & 4.199 & 19.36 & 6\\       
100413.14+630437.3 & 4.130 & 20.24 & 6 &       211450.34$-$063257.1 & 4.268 & 19.38 & 7\\     
104837.40$-$002813.6 & 4.031 & 19.05 & 6 &     214601.45$-$075343.6 & 4.192 & 19.43 & 7\\     
105254.59$-$000625.8 & 4.173 & 19.54 & 6 &     215817.60$-$010555.1 & 4.132 & 19.73 & 7\\     
105602.36+003222.0 & 4.064 & 19.66 & 6 &       221320.00+134832.5 & 4.129 & 19.78 & 7\\       
110819.16$-$005823.9 & 4.604 & 19.87 & 6 &     221705.72+135352.7 & 4.348 & 20.09 & 7\\       
112956.10$-$014212.3 & 4.850 & 19.64 & 7 &     222050.81+001959.0$^*$ & 4.700 & 20.21 & 6\\   
113745.66+012715.1 & 4.070 & 19.72 & 7 &       222807.58+003526.2 & 4.631 & 19.80 & 0\\       
115158.23+030341.7 & 4.701 & 20.47 & 7 &       223521.22$-$082127.2 & 4.372 & 20.07 & 7\\     
120640.73+033414.9 & 4.386 & 19.86 & 7 &       224740.17$-$091511.7 & 4.174 & 20.28 & 7\\     
123115.90$-$020506.0 & 4.144 & 19.75 & 7 &     224922.94$-$010745.8 & 4.003 & 19.71 & 0\\     
123347.21$-$014853.8 & 4.259 & 19.04 & 7 &     225246.44+142525.7 & 4.920 & 19.91 & 7\\       
123937.17+674020.7 & 4.425 & 20.18 & 6 &       230320.38$-$085433.1 & 4.337 & 20.16 & 7\\     
124757.44$-$011926.0 & 4.187 & 19.83 & 7 &     231010.59$-$100653.9 & 4.507 & 20.35 & 7\\     
125433.56$-$003922.7 & 4.291 & 20.10 & 6 &     232112.39+143312.0 & 4.026 & 20.24 & 7\\       
125802.61+022721.2 & 4.269 & 19.84 & 7 &       233255.72+141916.4 & 4.751 & 20.09 & 7\\       
130039.13+032203.8 & 4.155 & 20.09 & 7 &       235403.85+155630.3 & 4.573 & 20.34 & 7\\       
130216.13+003032.0 & 4.607 & 19.86 & 6 &       235718.36+004350.3 & 4.363 & 20.13 & 1\\       
132447.25$-$031358.2 & 4.052 & 19.79 & 6 &	& & & \\
\enddata
\tablecomments{See Table~1 for an explanation of the columns.
}
\end{deluxetable}

\end{document}